\documentclass[aps,preprint,epsfig,rotate]{revtex4}
\usepackage{graphicx}
\usepackage{bm}
\usepackage{epsfig}

\input epsf
\begin{document}
\title{Atomic few-body systems with muonium}

\author{Alexei M. Frolov}
\email[E--mail address: ]{afrolov@uwo.ca}

\affiliation{ITAMP, Harvard-Smithonian Center for Astrophysics, \\
         MS 14, 60 Garden Street, Cambridge MA 02138-1516, USA}  

\affiliation{Department of Applied Mathematics \\
       University of Western Ontario, London, Ontario N6H 5B7, Canada} 

\date{\today}

\begin{abstract}

Properties of some few-body systems which include one positively charged muon $\mu^{+}$ and two electrons $e^{-}$ are discussed. In particular, 
we consider the negatively charged muonium ion Mu$^{-}$ (or $\mu^{+} e^{-}_{2}$) and four-body MuPs (or $\mu^{+} e^{-}_{2} e^{+}$) systems 
each of which has only one stable bound (ground) state. The problem of annihilation of the electron-positron pair(s) in the MuPs system is 
investigated. The hyperfine structure splitting of the ground state in the MuPs system evaluated with our expectation value of the muon-positron 
delta-function is $\Delta \approx$ 23.05758 $MHz$. Another group of interesting four-body neutral systems investigated in this study includes 
the $p^{+} \mu^{+} e^{-}_2, d^{+} \mu^{+} e^{-}_2$ and $t^{+} \mu^{+} e^{-}_2$ `quasi-molecules'. These quasi-molecules are formed in 
large numbers when positively charged muons slow down in liquid hydrogen, or in liquid deuterium and/or tritium. The properties of these 
systems are unique, since they occupy an intermediate position between actual two-center molecules and one-center atoms.   

\noindent 
PACS number(s): 32.10.Fn, 31.15.A- and 31.15.Ve


\end{abstract}

\maketitle
\newpage

\section{Introduction}

In this communication we report the results of our analysis of some three- and four-body systems each of which include one positively charged muon $\mu^{+}$
and two electron $e^{-}$. Briefly, we can say that each of these few-body systems contains muonium Mu (or $\mu^{+} e^{-}$), or muonium ion Mu$^{-}$ (or 
$\mu^{+} e^{-}_2$). Recently, there is an increasing experimental interest to such few-body systems (see discussions and references in \cite{Mubook} - 
\cite{PRA16}). This can be explained by rapidly growing experimental abilities to detect and isolate similar few-particle systems. Moreover, by using the 
modern experimental techniques one can investigate some of the bound state properties of these systems. Another reason follows from the fact that the 
$\mu^{+}$ muon has positive electric charge and relatively small particle mass. This means that all atomic and molecular few-body systems which contain one 
positively charged muon and two electrons have very special electron density distribution which differs substantially from electron density distributions in 
`similar' atomic and molecular systems. In reality, such systems can be considered as a separate class of bound systems which are neither atoms, nor 
molecules. In some cases, the positively charged muon plays a role of central particle which stabilize a few-electron structure, e.g., in the Mu$^{-}$ ion 
and MuPs system (see below). Formally, all few-body systems with positively charged muon are unstable, but their life-time $\tau (\approx 2 \cdot 10^{-6}$ 
$sec$) significantly exceeds the mean time(s) of atomic transitions and decay processes $\tau_{tr} \le 1 \cdot 10^{-11}$ $sec$. In other words, these few-body 
quasi-atomic systems can be created and stabilized in their bound state(s) substantially faster than the decay of the $\mu^{+}$ muon can occur.    

As follows from the above, it is important to predict the overall stability of the few-body systems which include muonium Mu (or muonium ion Mu$^{-}$) and 
investigate their basic properties. These problems are considered in our study. In particular, below we consider the negatively charged Mu$^{-}$ (or $\mu^{+} 
e^{-}_{2}$) ion and neutral four-body MuPs (or $\mu^{+} e^{-}_{2} e^{+}$) systems. Each of these systems has only one bound (ground) state. Recently, we could
drastically improve the overall accuracy of our variational computations for the both Mu$^{-}$ ion and $MuPs$ system. For the ground $1S^{1}-$state in the 
Mu$^{-}$ ion we can now produce the variational wave functions which contain 2 - 3 additional (correct) decimal digits (compare with results from \cite{Fro2004}) 
in the total energy and other bound state properties. The overall accuracy of thr bound state computations for the Mu$^{-}$ ion is comparable with the accuracy
of the best atomic calculations for the two-electron atoms and ions. For the four-body MuPs system we have also derived much better variational approximations 
to the exact wave function. This allows us to evaluate many bound state properties of this system to very good accuracy (compare with \cite{Fro2015}). For the 
MuPs system we also re-derive the formulas which have been obtained and used earlier \cite{Fro2015} for some annihilation rates of the $(e^{+}, e^{-})-$pair.
Finally, a few corrections in such formulas have been made. 
           
A separate group of systems discussed below include three neutral $p^{+} \mu^{+} e^{-}_2, d^{+} \mu^{+} e^{-}_2$ and $t^{+} \mu^{+} e^{-}_2$ `quasi-molecules'. 
These four-body systems have never been considered earlier. Note that each of these three systems also contains muonium Mu (or $\mu^{+} e^{-}$), or negatively
charged muonium ion Mu$^{-}$ (or $\mu^{+} e^{-}_2$). The energy spectra and bound state properties of these four-body systems are substantially different from 
the spectra of the Mu$^{-}$ ion and MuPs systems. For instance, the bound state spectrum of each of the $p^{+} \mu^{+} e^{-}_2, d^{+} \mu^{+} e^{-}_2$ and 
$t^{+} \mu^{+} e^{-}_2$ quasi-molecules contains dozens of bound states which converge to the corresponding dissociation threshold(s). These three atomic systems 
are original, since they cannot be considered as the regular `two-center' molecules even in the first approximation. On the other hand, these systems are not
one-center atoms (or quasi-atoms). Formally, these three neutral $p^{+} \mu^{+} e^{-}_2, d^{+} \mu^{+} e^{-}_2$ and $t^{+} \mu^{+} e^{-}_2$ systems form a new
sub-class among all atomic and molecular systems.  

This paper has the following structure. The negatively charged muonium ion Mu$^{-}$ is considered in the next Section. Since this ion is a three-body system, 
then we can determine the total energy of the ground $1^1S-$state of the Mu$^{-}$ ion to very high numerical accuracy. Many expectation values (i.e. properties) 
of this ion in its ground state have also been determined in our highly accurate computations. The neutral four-body MuPs system is considered in the third 
Section. Here we also evaluate a number of few-photon annihilation rates for this system and the hyperfine structure splitting. The four-body $p^{+} \mu^{+} 
e^{-}_2, d^{+} \mu^{+} e^{-}_2$ and $t^{+} \mu^{+} e^{-}_2$ quasi-molecules are discussed in the fourth Section. Note that these systems have never been considered 
in earlier studies. For these three systems we consider only their electron ground states. Concluding remarks can be found in the last Section.

\section{Negatively charged muonium ion}

First, let us consider the negatively charged three-body muonium ion Mu$^{-}$ which consists of one positively charged muon $\mu^{+}$ and two electrons $e^{-}$. 
As mentioned above this system has only one stable bound state which is the ground $1^1S-$electron state. Stability of the ground state in the Mu$^{-}$ ion has 
been shown in 1986 \cite{Fro86}. A few years later this ion has been created in actual experiment \cite{Kuang}. The total energies and other bound state 
properties of this ion have been considered in \cite{Fro2004}. Since our paper \cite{Fro2004} was published the new experiments in high energy physics with 
positively charged muons have been conducted. The improved numerical value of the muonic mass now is:
\begin{eqnarray} 
  m_{\mu} = \frac{105.65836668 \; \; MeV/c^2}{0.510998910 \; \; MeV/c^2} = 206.76828191\ldots m_e  \; \; \; \label{eq1} 
\end{eqnarray}
The non-relativistic Hamiltonian of the Mu$^{-}$ ion takes the form
\begin{eqnarray}
 H = -\frac{\hbar^2}{2 m_e} \Bigl[\nabla^2_1 + \nabla^2_2 + \frac{m_e}{m_{\mu}} \nabla^2_3 \Bigr] - \frac{e^2}{r_{32}} - \frac{e^2}{r_{31}} + \frac{e^2}{r_{21}} 
 \; \; \; , \; \; \label{Hamil3}
\end{eqnarray}
where $\hbar = \frac{h}{2 \pi}$ is the reduced Planck constant, $m_e$ is the electron mass and $e$ is the absolute value of the electric charge of electron. In this 
equation the subscripts 1 and 2 designate the two electrons $e^-$, while the subscript 3 denotes the heavy central particle $\mu^{+}$ (quasi-nucleus) with the mass 
$m_{\mu}$ ($m_{\mu} \gg m_e$), and positive electric (nuclear) charge $+e$. For the Mu$^{-}$ ion we need to solve the non-relativistic Schr\"{o}dinger equation $H \Psi 
= E \Psi$ for the two-electron ions, where $H$ is the Hamiltonian, Eq.(\ref{Hamil3}), and $E (E < 0$) is the eigenvalue, which coinsides with the total energy of the 
Mu$^{-}$ ion in its ground $1^1S-$state. This state will be stable, if its total energy is lower than the total energy of the two-body muonium Mu ($\mu^{+} e^{-}$)
which equals $-\frac{m_{\mu}}{2 (m_{\mu} + 1)} = 0.5 (1 + m^{-1}_{\mu})$ in atomic units.  

Numerical solution of the Schr\"{o}dinger equation allows one to obtain highly accurate wave function(s) $\Psi$. In reality, the unknown (exact) wave function is 
approximated by using some rapidly convergent variational expansions. One of the best variational expansions known for three-body systems is the exponential variational 
expansion in the relative coordinates $r_{32}, r_{31}, r_{21}$, or in the perimetric coordinates $u_1, u_2, u_3$. For the ground (bound) $1^1S-$state of the two-electron 
ions/atoms the explicit form of these expansions are
\begin{eqnarray}
  \Psi &=& \frac12 ( 1 + \hat{P}_{12} ) \sum^{N}_{i=1} C_i \exp(-\alpha_i r_{32} - \beta_i r_{31} - \gamma_i r_{21}) \nonumber \\
  &=& \frac12 ( 1 + \hat{P}_{12} ) \sum^{N}_{i=1} C_i \exp[-(\alpha_{i} + \beta_{i}) u_{3} - (\alpha_{i} + \gamma_{i}) u_{2} - (\beta_{i} + \gamma_{i}) u_{3}] \; \; \; 
 \label{exp}
\end{eqnarray}
where $r_{ij} = \mid {\bf r}_i - {\bf r}_j \mid = r_{ji}$ are the interparticle scalar distances and ${\bf r}_i$ ($i$ = 1, 2, 3) are the three Cartesian coordinates of 
the particles. In Eq.(\ref{exp}) the notation $\hat{P}_{12}$ stands for the permutation operator of identical particles (electrons), $C_i$ ($i = 1, 2, \ldots, N$) are 
the linear parameters of the exponential expansion, Eq.(\ref{exp}), while $\alpha_i, \beta_i$ and $\gamma_i$ are the non-linear parameters of this expansion. These 
non-linear parameters must be varied in calculations to increase the overall efficiency and accuracy of the method. The best-to-date optimization strategy for these 
non-linear parameters was described in \cite{Fro2001}, while its modified (advanced) version was presented in another paper published in 2006 (see the second paper in 
Ref.\cite{Fro2001}). 

Note that the $3 N$ following conditions $\alpha_{i} + \beta_{i} > 0, \alpha_{i} + \gamma_{i} > 0, \beta_{i} + \gamma_{i} > 0$ (for $i = 1, 2, \ldots, N$) must always 
be obeyed for the non-linear parameters in Eq.(\ref{exp}) to guarantee convergence of all three-particle integrals which are needed in computations. Three scalar 
perimetric coordinates $u_i$ are simply related with the relative coordinates $r_{kj}$ by the following linear relations
\begin{eqnarray}
  & & u_1 = \frac12 ( r_{21} + r_{31} - r_{32}) \; \; \; , \; \; \; r_{32} = u_2 + u_3 \nonumber \\
  & & u_2 = \frac12 ( r_{21} + r_{32} - r_{31}) \; \; \; , \; \; \; r_{31} = u_1 + u_3 \; \; \; \label{coord} \\
  & & u_3 = \frac12 ( r_{31} + r_{32} - r_{21}) \; \; \; , \; \; \; r_{21} = u_1 + u_2 \nonumber
\end{eqnarray}
where $r_{ij} = r_{ji}$. In contrast with the relative coordinates $r_{32}, r_{31}, r_{21}$ the three perimetric coordinates $u_1, u_2, u_3$ are independent of each other 
and each of them varies between 0 and $+\infty$. This drastically simplifies analytical and numerical computations of all three-body integrals which are needed for solution 
of the corresponding eigenvalue problem and for evaluation a large number of bound state properties of the Mu$^{-}$ ion. Additional interesting details about perimetric 
coordinates can be found in the Appendix.      
 
The total energies and bound state properties of this ion (in atomic units) can be found in Tables I (energies) and II (properties). All these values have been determined 
to very high numerical accuracy with the use of the extended precision arithmetic \cite{Bail1}. Analogous bound state properties of the Mu$^{-}$ ion are slightly different 
from values presented in \cite{Fro2004} which can be explained by the differences in the muonium mass $m_{\mu}$. It should be mentioned that the current accuracy of 
numerical computations for three-particle (non-relativistic) Coulomb systems is extremely high and this allows one to determine the energies and wave functions of all bound 
states to very high accuracy which can be evaluated as $\approx 1 \cdot 10^{-22} - 1 \cdot 10^{-25}$ $a.u.$ The overall accuracy of our numerical calculations of the ground
$1^{1}S-$state in the Mu$^{-}$ ion $\approx 3 \cdot 10^{-24}$ $a.u.$ is quite comparable to the accuracy of the best atomic bound state computations for two-electron atoms 
and ions. Highly accurate wave functions can be used to obtain the correct numerical values of many bound state properties, including a few slowly convergent properties 
\cite{FroWa16}, e.g., the $\langle r^{6}_{ij} \rangle, \langle r^{8}_{ij} \rangle$ and $\langle r^{10}_{ij} \rangle$  expectation values. As expected the numerical values 
of all computed bound state properties of the Mu$^{-}$ ion are close to the analogous properties of the negatively charged hydrogen ion H$^{-}$ which are also shown in Table 
II (see \cite{Fro2015}). The results from Table I and II is the most complete and accurate set of the expectation values ever determined for the muonium ion Mu$^{-}$.    
   
The negatively charged muon Mu$^{-}$ ion is an example of three-body quasi-atomic system where the $\mu^{+}$ muon plays a role of a central particle, or `quasi-nucleus'. 
Another such a system with the same quasi-nucleus is the four-body neutral system MuPs (or $\mu^{+} e^{-}_2 e^{+}$) which is discussed below. Note that the MuPs system has a 
more complicated internal structure and a number of interesting bound state properties.   

\section{Four-body muonium-positronium system}

Another systems which contains muonium Mu ($\mu^{+} e^{-}$), or muonium ion Mu$^{-}$ ($\mu^{+} e^{-}_2$) is the four-body MuPs (or $\mu^{+} e^{-}_2 e^{+})$ quasi-atom which 
contains one bound positron \cite{Fro2015}. In other words, the MuPs `quasi-atom' is a pure leptonic four-body quasi-atom which also contains one positively charged muon 
(heavy `central' particle), two electrons and one positrons (light particles). Such a unique system can now be created and observed in actual experiments in which a large 
number of positively charged muons are created. The stability of the MuPs system follows from the fact that the analogous four-body HPs and Ps$_2$ systems are stable. 
Moreover, since each of the HPs and Ps$_2$ systems has only one bound state (ground state), then we conclude that the analogous MuPs systems also has only one bound (ground) 
state which is the $1^1S-$electron state. In atomic units the non-relativistic Hamiltonian of the four-body MuPs system takes the form (see, e.g., \cite{LLQ})
\begin{eqnarray}
 H = -\frac{1}{2} \Bigl[\nabla^2_1 + \nabla^2_2 + \nabla^2_3 + \frac{1}{m_{\mu}} \nabla^2_4 \Bigr] - \frac{1}{r_{32}} - \frac{1}{r_{31}} - \frac{1}{r_{42}} 
 - \frac{1}{r_{41}} + \frac{1}{r_{21}} + \frac{1}{r_{43}} \; \; \; , \; \; \label{Hamil4}
\end{eqnarray}
where the particles 1 and 2 are the two electrons $e^{-}$ (or `-'), while the particles 3 and 4 are the positron $e^{+}$ (or `+') and muon $\mu^{+}$, respectively. The MuPs 
system contains two light negative particles and one light positive particle positron $e^{+}$ which are moving around one heavy center $\mu^{+}$. In this sense the MuPs system 
is similar to the four-body HPs system(s). The mass of the positively charged muon used n this Section is $m_{\mu} = 206.768264 m_e$. 

To determine the accurate solution(s) of the non-relativistic Schr\"{o}dinger equation $H \Psi = E \Psi$ in this Section we apply the variational expansion of the wave function 
written in multi-dimensional gaussoids, which are, in fact, the six-dimensional gaussoids for the four-body MuPs system. Each of the basis functions in this expansion explicitlty 
depends upon all six relative coordinates $r_{ij}$, where $r_{ij} = r_{12}, r_{13}, r_{14}, r_{23}, r_{24}$ and $r_{34}$. For the ground state of the MuPs system the variational 
expansion in six-dimensional gaussoids takes the form (see, e.g., \cite{KT}, \cite{Fro2010}):
\begin{eqnarray}
  \psi(L = 0; S = 0) = \frac{1}{2} (1 + \hat{P}_{12}) \sum^{N_A}_{i=1} C_i \exp(-a_{ij} r^2_{ij}) \chi_{S=0} \; \; \; \label{gaus1}
\end{eqnarray}
where $C_i$ are the linear variational coefficients of the variational function, while $a_{ij}$,  where $(ij)$ = (12), (13), $\ldots$, (34), are the six non-linear parameters 
in the radial function associated with the $\chi_{S=0}$ spin function. This spin function can be chosen in the form 
\begin{eqnarray}
  \chi_{S=0} = (\alpha_1 \beta_2 - \beta_1 \alpha_2) \alpha_{\mu} \alpha_{+} = \alpha_{4} \alpha_{3} (\alpha_{1} \beta_{2} - \beta_{1} \alpha_{2}) \; \; \; \label{spin}
\end{eqnarray}
where the notations $\alpha$ and $\beta$ are used spin-up an spin-down functions (see, e.g., \cite{LLQ}). The total energies and bound state properties of the MuPs system (in 
atomic units) determined with the use of the trial functions, Eq.(\ref{gaus1}), can be found in Table III. In general, the current numerical computations of the bound states in 
the four-body systems with comparable particle masses is singificantly less than such an accuracy obtained in calculations of the Coulomb three-body systems (see, e.g., the 
Mu$^{-}$ ion discussed above). This can be explained by a different differential geometry of the general four-body problem. Briefly, we can say that follows from the fact that 
it is impossible to define six perimetric coordinates (for four-body systems) which have the same properties as the three perimetric coordinates introduced above for three-body 
systems (see Appendix). 

The expectation values from Table III can be considered as `elementary' properties. However, by using these `elementary' properties we can determine (or evaluate) some actual 
properties of the MuPs system. In particular, by using the expectation value of the electron-positron delta-function from Table III we can evaluate the two- and three-photon 
annihilation rates $\Gamma_{2 \gamma}$ and $\Gamma_{3 \gamma}$ for the ground state in the MuPs system. The sum of these two annihilation rates essentially determines the life-time 
of the MuPs system. The formula for the $\Gamma_{2 \gamma}$ rate takes the form \cite{Fro2015}
\begin{eqnarray}
 \Gamma_{2 \gamma}({\rm MuPs}) = 2 \pi \alpha^4 c a^{-1}_0 \langle \delta({\bf r}_{+-}) \rangle = 100.939408683 \cdot 10^{9} \langle \delta({\bf r}_{+-}) \rangle \; sec^{-1} 
 \; \; \; . \label{An2g}
\end{eqnarray}
where $\langle \delta_{+-} \rangle$ is the expectation value of the electron-positron delta-function determined for the ground (bound) $1^{1}S_{e}-$state in the MuPs system. 
The indexes `+' and `-' used in delta-functions in Eq.(\ref{An2g}) designate the positron and electron, respectively, while the index $\mu$ stands for the $\mu^{+}$ 
muon. In Eq.(\ref{An2g}) and everywhere below the notation $\alpha = \frac{e^2}{\hbar c} = 7.2973525698 \cdot 10^{-3} \Bigl(\approx \frac{1}{137}\Bigr)$ is the dimensionless 
fine-structure constant, $c = 2.99792458 \cdot 10^{8}$ $m \cdot sec^{-1}$ is the speed of light in vacuum, and $a_0$ is the Bohr radius which equals $0.52917721092 \cdot 
10^{-10}$ $m$ \cite{CRC}. Analogous formula for the three-photon annihilation rate $\Gamma_{3 \gamma}({\rm MuPs})$ takes the form 
\begin{eqnarray}
 \Gamma_{3 \gamma}({\rm MuPs}) = 2 \frac{4 (\pi^2 - 9)}{3} \alpha^5 c a^{-1}_0 \langle \delta({\bf r}_{+-}) \rangle = 2.718545954 \cdot
 10^{8} \langle \delta({\bf r}_{+-}) \rangle \; sec^{-1} \label{An3g}
\end{eqnarray}
This formula can easily be derived from the well known formula obtained in \cite{Ore} for one triplet electron-positron pair. The lowest-order QED correction to the 
$\Gamma_{2 \gamma}({\rm MuPs})$ annihilation rate is written in the form (derived from the formula \cite{PR1957} for one singlet electron-positron pair)
\begin{eqnarray}
 \Delta \Gamma_{2 \gamma}({\rm MuPs}) = - 2 \pi \alpha^5 c a^{-1}_0 \Bigl( \frac{13 \pi}{12} - \frac{7}{\pi} \Bigr) 
 \langle \delta({\bf r}_{+-}) \rangle = - 8.6565792289 \cdot 10^{8} \langle \delta({\bf r}_{+-}) \rangle \; sec^{-1} 
 \; \; \; . \label{An2gc}
\end{eqnarray} 
This correction is negative. By using these formulas and our expectation value for the electron-positron delta-function $\langle \delta({\bf r}_{+-}) \rangle$ from Table III we 
have found that $\Gamma_{2 \gamma}$(MuPs) = 2.464793493$\cdot 10^{9}$ $sec^{-1}$, $\Gamma_{3 \gamma}$(MuPs) = 6.63829367$\cdot 10^{6}$ $sec^{-1}$ and 
$\Delta \Gamma_{2 \gamma}({\rm MuPs})$ = - 2.1138106945 $\cdot 10^{7}$ $sec^{-1}$, respectively. Numerical values of the four- and five-photon annihilation rates 
($\Gamma_{4 \gamma}$ and $\Gamma_{5 \gamma}$) are obtained from the $\Gamma_{2 \gamma}({\rm MuPs})$ and $\Gamma_{3 \gamma}({\rm MuPs})$ rates mentioned above with the use of 
the following formulas (derived from the original formulas obtained in \cite{PRA1983})
\begin{eqnarray}
 \Gamma_{4 \gamma}({\rm MuPs}) = 0.274 \Bigl(\frac{\alpha}{\pi}\Bigr)^2 \Gamma_{2 \gamma}({\rm MuPs}) \; \; \; , \; \; \;  
 \Gamma_{5 \gamma}({\rm MuPs}) = 0.177 \Bigl(\frac{\alpha}{\pi}\Bigr)^2 \Gamma_{3 \gamma}({\rm MuPs})
 \; \; \; . \label{An45g}
\end{eqnarray} 
With our expectation values of the electron-positron delta-function $\delta({\bf r}_{+-})$ (see Table III) are 3.64386$\cdot 10^{3}$ $sec^{-1}$ and 6.33958 $sec^{-1}$. 

The one-photon positron annihilation in the MuPs system is of interest in some problems as well as for the future development of QED. This process can be represented as a regular 
two-photon annihilation of the $(e^{-}, e^{+})-$pair and the following absorbtion of one of the emitted photons either by the second electron, or by the $\mu^{+}$ muon (heavy 
particle). Therefore, one can observe, in principle, the two one-photon annihilations in MuPs. The rates of these processes are designated below as 
$\Gamma^{(e)}_{1 \gamma}({\rm MuPs})$ and $\Gamma^{(\mu)}_{1 \gamma}({\rm MuPs})$. Analytical expressions for these rates can be derived from the results of earlier studies
\cite{Kryu} and \cite{Fermi}. In atomic units the explicit formulas for the $\Gamma^{(e)}_{1 \gamma}({\rm MuPs})$ and $\Gamma^{(\mu)}_{1 \gamma}({\rm MuPs})$ annihilation rates 
take the form
\begin{eqnarray}
 \Gamma^{(e)}_{1 \gamma}({\rm MuPs}) = \frac{64 \pi^2}{27} \alpha^8 c a^{-1}_0 \langle \delta({\bf r}_{+--}) \rangle = 1065.75691955  \langle \delta({\bf r}_{+--}) \rangle 
 \; \; \; , \label{An1ge}
\end{eqnarray} 
and 
\begin{eqnarray}
 \Gamma^{(\mu)}_{1 \gamma}({\rm MuPs}) = \frac{4 \pi^2}{3} \alpha^8 Q^5 c a^{-1}_0 \langle \delta({\bf r}_{+-\mu}) \rangle = 599.48826725 \langle \delta({\bf r}_{+-\mu}) \rangle 
 \; \; \; , \label{An1gmu}
\end{eqnarray} 
where $Q = 1$ for MuPs, while $\langle \delta({\bf r}_{+--}) \rangle$ and $\langle \delta({\bf r}_{+-\mu}) \rangle$ are the expectation values of the corresponding triple 
delta-functions determined in atomic units. In derivation of the formula, Eq.(\ref{An1gmu}), we have assumed that the muonium mass is very large (infinite). The sum of these values 
allows one to determine the one-photon annihilation rate for the MuPs system
\begin{eqnarray}
 \Gamma_{1 \gamma}({\rm MuPs}) = \frac{4 \pi^2}{3} \alpha^8 c a^{-1}_0 \Bigl[ \langle \delta({\bf r}_{+-\mu}) \rangle + \frac{16}{9} \langle \delta({\bf r}_{+--}) \rangle \Bigr] 
  \; \; \; , \label{An1g}
\end{eqnarray} 
By using our expectation values of the triple delta-functions mentioned in these equations (see Table III) we have found the following numerical values 
$\Gamma^{(e)}_{1 \gamma}({\rm MuPs}) \approx$ 0.39087487 $sec^{-1}$, $\Gamma^{(\mu)}_{1 \gamma}({\rm MuPs}) \approx$ 0.51148759 $sec^{-1}$ and $\Gamma_{1 \gamma}({\rm MuPs}) 
\approx$ 0.90236246 $sec^{-1}$.

We can also evaluate the hyperfine structure splitting in the MuPs system, where the hyperfine structure arises from direct interaction between the spin-vectors of the positron 
and muon. The hyperfine structure splitting in the MuPs system $(\Delta H)_{hsp}$ is written in the form \cite{Fro2015} 
\begin{equation}
 (\Delta H)_{hsp} = \frac{8 \pi \alpha^2}{3} \mu^2_B \frac{g_{\mu}}{m_{\mu}} \frac{g_{e}}{m_{e}} \; \langle \delta({\bf r}_{+ \mu}) \rangle = 14229.1255 \; 
 \langle \delta({\bf r}_{+ \mu}) \rangle \label{spl31}
\end{equation}
where $\alpha$ is the fine-structure constant, $\mu_B$ is the Bohr magneton which equals 0.5 in atomic units, while $\langle \delta_{+ \mu} \rangle$ is the expectation value of 
the muon-positron delta-function. Also in Eq.(\ref{spl31}) the notations $m_{\mu}$ and $m_{e}$ stand for the muon/positron masses-at-rest, while the factors $g_{+}$ = 
-2.0023193043718 and $g_{\mu}$ = -2.0023318396 are the gyromagnetic ratios. In Eq.(\ref{spl31}) we have used the factor 6.579 683 920 61$\cdot 10^9$ ($MHz/a.u.$) has been used 
to re-calculate the $\Delta E_{hf}$ energy from atomic units to $MegaHertz$. By using the expectation value of the muon-positron delta-function $\langle \delta_{+ \mu} \rangle 
\approx 1.62045 \cdot 10^{-3}$ $a.u.$ from Table III, one finds that the value $a$ in Eq.(\ref{spl31}) equals $(\Delta H)_{hsp} \approx 23.05758$ $MHz$. This coincides with the 
energy difference between the hyperfine structure states with $J = 0$ and $J = 1$, where the notation $J$ stands for the total spin of the muon-positron pair in the MuPs system.

To conclude this Section let us note that there is another interesting process which can be observed, in principle, in the MuPs system. This process is the conversion of the 
$\mu^{+} - e^{-}$ pair into its charge-conjugate $\mu^{-} - e^{+}$ pair. During the original MuPs system is transformed into another (new) Mu$^{*}$Ps system which is also bound 
and contains two positrons. The overall probability of this process is very small, but it is proportional to the $\langle \delta({\bf r}_{- \mu}) \rangle$ expectation value (see 
Table III).

\section{Four-body hydrogen-muonium quasi-molecules}

The third group of few-body systems with positively charged muons considered in this study includes three four-body `quasi-molecules': $p^{+} \mu^{+} e^{-}_2, d^{+} \mu^{+} 
e^{-}_2$ and $t^{+} \mu^{+} e^{-}_2$. These systems are always formed when high-energy positively charged muons slow down in the liquid hydrogen, or in liquid deuterium and/or 
tritium. Each of these quasi-molecules includes one positively charged muon and two bound electrons. In each of these three quasi-molecules the $\mu^{+}$ muon plays the role 
of the second positively charged `heavy' center. Indeed, as follows from the mass ratio of the muon and electron masses the muon's velocity in the $a \mu e_2$ (= 
$a^{+} \mu^{+} e^{-}_2$) system is substantially smaller ($\sim 206.768$ times smaller) than the electron's velocity. This means that in the first approximation the $\mu^{+}$ 
muon can be considered as the second heavy particle in the $a \mu e_2$ quasi-molecule, where $a = p, d, t$. However, the actual small dimensionless parameters in the 
Born-Oppenheimer theory \cite{BO} (see also \cite{Beth65}) of 
the adiabatic two-center systems are defined as follows (in our notations):
\begin{eqnarray}
  \tau_1 = \sqrt[4]{\frac{M_a}{M_a + m_{\mu} + 2 m_e}} \; \; \;  and \; \; \; \tau_2 = \sqrt[4]{\frac{m_{\mu}}{M_a + m_{\mu} + 2 m_e}} \label{mass} 
\end{eqnarray}
where $a = (p, d, t)$ and notations $m_{\mu}$ and $M_a$ stand for the masses of the two heavy particles. In our calculations we used the following numerical values for the
particle masses: $M_p = 1836.152701 m_e, M_d = 3670.483014 m_e, M_t = 5496.92158 m_e$ and $m_{\mu} = 206.768262 m_e$. The two dimensionless parameters $\tau_1$ and $\tau_2$ 
determine those systems to which adiabatic approximation can be applied. Formally, the adiabatic approximation works well in those cases when $\tau_{min} = \min (\tau_1, \tau_2) 
\ge 0.5$. For the $p^{+} \mu^{+} e^{-}_2, d^{+} \mu^{+} e^{-}_2$ and $t^{+} \mu^{+} e^{-}_2$ quasi-molecules the parameter $\tau_{min}$ does not exceed 0.15, i.e. it is 
significantly smaller than 0.5. This means that any of the three $a \mu e_2$ systems investigated in this study cannot be considered as an actual molecule with the two heavy 
centers (or two-center molecule). The main reason for this is obvious, since the directions of the central $a^{+} \leftrightarrow \mu^{+}$ axis is not fixed in space even in the 
$t \mu e_2$ system. In other words, the mass of the positively charged $\mu^{+}$ `center' is too small to consider this particle as an actual `second' heavy center in the 
$a \mu e_2$ system. In reality, the $p^{+} \mu^{+} e^{-}_2, d^{+} \mu^{+} e^{-}_2$ and $t^{+} \mu^{+} e^{-}_2$ quasi-molecules form a separate class of quasi-atomic systems which 
differs substantially from regular atoms and molecules. Below, we consider these three four-body systems as `quasi-molecules', but this name is confusing, since it does not 
represent the actual situation with these systems.    

In atomic units the non-relativistic Hamiltonian of the four-body $a \mu e_2$ systems is written in the form
\begin{eqnarray}
 H = -\frac{1}{2} \Bigl[\nabla^2_1 + \nabla^2_2 + \frac{1}{m_{\mu}} \nabla^2_3 + \frac{1}{M_a} \nabla^2_4 \Bigr] - \frac{e^2}{r_{32}} - \frac{e^2}{r_{31}} - \frac{e^2}{r_{42}} 
 - \frac{e^2}{r_{41}} + \frac{e^2}{r_{21}} + \frac{e^2}{r_{43}} \; \; \; , \; \; \label{Hamil45}
\end{eqnarray}
where $e = 1$ and the particles 1 and 2 are the two electrons, while the particles 3 and 4 are the positively charged muon and heavy particle $a$, where $ a = p, d, t$, 
respectively. In Eq.(\ref{Hamil45}) we assume that the masses of two heavy particles $a^{+}$ and $\mu^{+}$ are expressed in terms of the electron mass $m_e$. The $a \mu e_2$ 
system contains two light negative particles (electrons) and two heavy positive particles (or two centers) which form the `molecular' axis $a^{+} \leftrightarrow \mu^{+}$. For 
pure adiabatic two-center molecule the location of this axis in space must be fixed, but this is not the case for the $p^{+} \mu^{+} e^{-}_2, d^{+} \mu^{+} e^{-}_2$ and 
$t^{+} \mu^{+} e^{-}_2$ quasi-molecules. 

In general, the properties of the $a \mu e_2$ quasi-molecular system must be similar to the properties of the two-center hydrogen molecules such as D$_2$, DT and/or HD. Direct 
comparison of the numerical values of bound state properties determined for the $p^{+} \mu^{+} e^{-}_2, d^{+} \mu^{+} e^{-}_2$ and $t^{+} \mu^{+} e^{-}_2$ quasi-molecules with 
the analogous properties of the H$_2$ molecule indicates clearly that such a similarity is observed. However, the overall accuracy of such a similarity is not high and cannot 
be improved, e.g., by using more accurate wave functions. In reality, the $p \mu e_2, d \mu e_2$ and $t \mu e_2$ quasi-molecular systems form a separate class among four-body 
systems with two electrons (see above). The same conclusion has been mentioned in \cite{Fro2004}, where we discussed the analogous three-body ions: $(p \mu e)^{+}, (d \mu e)^{+}$ 
and $(t \mu e)^{+}$ (see also \cite{Mac1} and \cite{Mac2}).  

In order to solve the non-relativistic Schr\"{o}dinger equation $H \Psi = E \Psi$ for the $a^{+} \mu^{+} e^{-}_2$ systems we applied the same variational expansion in 
multi-dimensional gaussoids, Eq.(\ref{gaus1}). The non-linear parameters $a_{ij}$ in this expansion, Eq(\ref{gaus1}), must carefully be optimized to produce accurate numerical 
results, including the total energies of these systems. By performing a number of consequtive optimizations of these parameters we have produced a number of trial wave 
functions which can be considered as accurate approximations to the actual wave functions of these four-body systems. A large number of bound state properties of the 
$p^{+} \mu^{+} e^{-}_2, d^{+} \mu^{+} e^{-}_2$ and $t^{+} \mu^{+} e^{-}_2$ quasi-molecules in their ground state(s) can be found in Table III. By using the expectation values 
from Table III one can predict numerical values of some actual properties, i.e properties which can be observed in experiments. For instance, by using the expectation values of 
the muon-proton (or muon-deutron, etc) delta-function ($\langle \delta({\bf r}_{a \mu}) \rangle$ in our notations) we can evaluate the hyperfine structure splitting in each of 
the $a \mu e_2$ considered in this study.

The hyperfine structure splitting in the ground $S-$states of the four-body $p^{+} \mu^{+} e^{-}_2, d^{+} \mu^{+} e^{-}_2$ and $t^{+} \mu^{+} e^{-}_2$ quasi-molecules is the
result of the direct spin-spin interaction between muonic spin and spin of the heavy hydrogen nucleus, i.e. protium, deuterium and tritium. The two electrons form a singlet 
pair with zero contribution into the hyperfine structure. The formula which is used to evaluate the hyperfine structure splitting $(\Delta H)_{hsp}$ in the ground $S-$states of 
each of the four-body $a \mu e_2$ quasi-molecule takes the form (in atomic units)
\begin{eqnarray}
  (\Delta H)_{hsp} &=& \frac{2 \pi}{3} \alpha^2 \frac{g_N g_{\mu}}{m_p m_{\mu}} \langle \delta({\bf r}_{\mu N}) \rangle ({\bf I}_N \cdot {\bf s}_{\mu}) 
   = \frac{\pi}{3} \alpha^2 \frac{g_N g_{\mu}}{m_p m_{\mu}} \langle \delta({\bf r}_{\mu N}) \rangle [ J (J + 1) \nonumber \\ 
 &-& I_N (I_N + 1) - s_{\mu} (s_{\mu} + 1) ] \; \; \; , \; \; \label{spl1} 
\end{eqnarray} 
where $\langle \delta({\bf r}_{\mu N}) \rangle$ is the expectation value of the muon-nucleus delta-function, while other notations have exactly the same meaning and numerical
values as in Eq.(\ref{spl31}). Note that in atomic units the Bohr magneton $\mu_B$ exactly equals $\frac12$. The factor $g_{N}$ in Eq.(\ref{spl31}) equals to the ratio 
$\frac{\cal M}{I_N}$, where ${\cal M}_{p} = 2.792847386, {\cal M}_{d} = 0.857438230$ and ${\cal M}_{t} = 2.9789624775$, while $I_{p} = \frac12, I_{d} = 1$ and $I_{t} = \frac12$. 
The factor $m_p$ = 1836.152701 $me$ is the proton's mass expressed in the electron mass $m_e$. Also, in Eq.(\ref{spl1}) the factor 6.579 683 920 61$\cdot 10^9$ ($MHz/a.u.$) must 
been used to re-calculate the $\Delta E_{hf}$ energy from atomic units to $MegaHertz$.  By using these numerical values we can evaluate the coefficients which arise in 
Eq.(\ref{spl1}) for the $p \mu e_2, d \mu e_2$ and $t \mu e_2$ quasi-molecules. These coefficients are: -10.808930550, -1.651657610 and -11.529236682, respectively. Now, by 
using the expectation values of the muon-nucleus delta-functions from Table III we obtain the following hyperfine structure for these four-body systems: (a) $\varepsilon(J = 1)$ 
= -30.098 $Hz$ and $\varepsilon(J = 0)$ = 90.295 $Hz$ for $p \mu e_2$, (b) $\varepsilon(J = \frac32)$ = -12.567 $Hz$ and $\varepsilon(J = \frac12)$ = 25.135 $Hz$ for $d \mu e_2$, 
and (c) $\varepsilon(J = 1)$ = -46.023 $Hz$ and $\varepsilon(J = 0)$ = 138.070 $Hz$ for $t \mu e_2$. The differences between these two energies gives the corresponding hyperfine 
structure splitting for each quasi-molecule (120.393 $Hz$, 37.702 $Hz$ and 184.093 $Hz$, respectively). Very small values of these hyperfine structure splittings indicate that each 
of these system has internal structure which is close to the adiabatic two-center molecule. On the other hand, these numerical values are in dozens times larger than values which 
can be found in any actual two center molecule.   
 
\section{Conclusion}
 
We have considered a number of systems which contain the muonium Mu, or muonium ion Mu$^{-}$. In other words, these systems include one positively charged muon $\mu^{+}$ and two 
boound electrons $e^{-}$. The total energies and a large number of bound state properties of the negatively charged muonium ion Mu$^{-}$ have been obtained from highly accurate 
numerical computations. The ground state in the four-body MuPs (or $\mu^{+} e^{-}_{2} e^{+}$) system is also studied in detail. It is shown that the bound state properties are 
similar to analogous properties of the HPs four-body system. For MuPs we evaluate the two- and three-photon annihilation rates and hyperfine structure splitting. The computed 
expectation values agree very well with the results of earlier studies (see, e.g., \cite{Fro2015}), but our current results have significantly better numerical accuracy. 

In this study we also investigate the bound (ground) states in the four-body quasi-molecules $p^{+} \mu^{+} e^{-}_2, d^{+} \mu^{+} e^{-}_2$ and $t^{+} \mu^{+} e^{-}_2$ (or 
$p \mu e_2, d \mu e_2$ and $t \mu e_2$). These systems have never been considered in earlier studies. The three quasi-molecules $p \mu e_2, d \mu e_2$ and $t \mu e_2$ cannot be 
considered as the `pure adiabatic' two-center molecules, since the positively charged muon $\mu^{+}$ is a relatively light particle. This fact allows us to apply some regular 
methods, e.g., the variational expansion, Eq.(\ref{gaus1}), to study the internal structure of these `intermediate' systems. If the $p^{+} \mu^{+} e^{-}_2, d^{+} \mu^{+} e^{-}_2$ 
and $t^{+} \mu^{+} e^{-}_2$ would be truly adiabatic (or two-center) systems, then the expansion, Eq.(\ref{gaus1}), could not be used directly, i.e. without modifications, to 
produce highly accurate results for such systems. By using our `regular' variational expansion in six-dimensional gaussoids we have determined a number of bound state properties 
of the ground state(s) in these four-body quasi-molecules. In addition to this we investigated the hyperfine structure and evaluated the hyperfine structure splitting in each of 
the $p^{+} \mu^{+} e^{-}_2, d^{+} \mu^{+} e^{-}_2$ and $t^{+} \mu^{+} e^{-}_2$ systems. 

\section{Appendix. Perimetric coordinates}

Note that the three-body perimetric coordinates $u_1, u_2, u_3$ were known to antic greeks. The Heron's formula (derived first by Hero of Alexandria, but very likely it was also 
known to Archimedes 200 years earlier) gives the area of a triangle $S$ by whose sides have lengths $r_{32}, r_{31}$ and $r_{21}$:
\begin{eqnarray}
  S = \sqrt{p u_1 u_2 u_3} \; \; \; \label{Aeq1} 
\end{eqnarray} 
where $u_1, u_2, u_3$ are the perimetric coordinates defined above (see, Eq.(\ref{coord})), while $p = \frac12 (u_1 + u_2 + u_3) = \frac12 (r_{32} + r_{31} + r_{21})$ is the 
semi-perimeter of this triangle which coincides with the half of the sum of three perimetric coordinates (and three relative coordinates). Here we assume that the indexies 1, 
2 and 3 stand for the three vertexes of the triangle. To transform this formula into the actual three-body problem we need to place three point particles in three vertexes of
of this triangle. Simplicity of the formula, Eq.(\ref{Aeq1}), is the first indication of high efficiency of the perimetric coordinates in application to the analysis of various 
three-body problems. Three perimetric coordinates have been introduced in actual numerical computations of three-body systems by C.L. Pekeris almost 60 years ago \cite{Pek1}, 
\cite{Pek2}. The use of these coordinates was crucial to obtain highly accurate solutions for a large number of three-body problems. It is interesting to note that C.L. Pekeris 
always considered himself as a pure `geophysicist' which has a restricted interest in atomic physics (a `personal hobby' - according to Yu.N. Demkov who met with Pekeris in the 
early 1970's). Now a large mehods based on the use of the perimetric coordinates are extensively used in various three-body problems and, in particular, for analytical and 
numerical computations of different three-body integrals, including very complex and singular integrals. 

An obvious success of the perimetric coordinates for various three-body problems stimulated discussion about the four-body perimetric coordinates which can be used for highly
accurate solutions of different four-body problems \cite{Fro2006A}. Indeed, for an arbitrary four-body system we can introduce twelve perimetric coordinates which must obey six 
additional conditions (or constraints) \cite{Fro2006A}. Formally, we can exclude six (of twelve) perimetric coordinates by solving these six constraints as algebraic equations. 
The six remaining perimetric coordinates will be sufficient for a complete description of an arbitrary four-body problem. However, during such a procedure some useful properties 
of the perimetric coordinates can be lost, e.g., some of them can be negative, or vary between the lower and upper limits $A$ and $B$, where $A \ne 0$ and $B \ne +\infty$. In 
addition to this, we have a problem of re-ordering of the arising perimetric coordinates (this problem does not exist for three-body systems). This drastically complicates 
analytical and numerical computations of actual four-body systems with the use of the four-body perimetric coordinates. Note also that the problem of the correct definition of 
the four-body perimetric coordinates has been re-investigated recently \cite{student}.            
 
\section{Acknowledgments}

This work was supported in part by the NSF through a grant for the Institute for Theoretical Atomic, Molecular, and Optical Physics (ITAMP) at Harvard University and 
the Smithsonian Astrophysical Observatory. Also, I wish to thank James Babb (ITAMP) and David M. Wardlaw for stimulating discussion.

\newpage
\newpage
\begin{table}[tbp]
   \caption{Convergence of the total energies $E$ (in $a.u.$) determined for the ground $1^{1}S-$state of the Mu$^{-}$ ion. The notation $N$ is the total 
            number of basis functions used.}
     \begin{center}
     \begin{tabular}{| l | l | l |}
      \hline\hline
 $N$  & $E$ (series A) & $E$ (series B) \\ 
     \hline
 400$^{(a)}$  &  -0.525054806501688          & -0.525054806501688      \\
        \hline
 3300         &  -0.525054806501730774024348 & -0.52505480650173077402418 \\

 3500         &  -0.525054806501730774025187 & -0.52505480650173077402500 \\

 3700         &  -0.525054806501730774025663 & -0.52505480650173077402547 \\

 3840         &  -0.525054806501730774025914 & -0.52505480650173077402573 \\ 
         \hline\hline
  \end{tabular}
  \end{center}
$^{(a)}$The short-term cluster wave function with the carefully optimized non-linear parameters.  
  \end{table}
%

%
\begin{table}[tbp]
   \caption{The expectation values of some propeties (in atomic units $a.u.$) for the Mu$^{-}$ and ${}^{\infty}$H$^{-}$ ions.}
     \begin{center}
     \scalebox{0.90}{%
     \begin{tabular}{| c | c | c | c | c |}
      \hline\hline
 ion & $\langle r^{-2}_{eN} \rangle$ & $\langle r^{-2}_{ee} \rangle$ & $\langle r^{-1}_{eN} \rangle$ & $\langle r^{-1}_{ee} \rangle$ \\
      \hline 
             Mu$^{-}$ & 1.10551221416860383 & 0.15325557249058329 & 0.6796545010765918638 & 0.309199389149722180 \\ 

 ${}^{\infty}$H$^{-}$ & 1.11666282452542572 & 0.15510415256242466 & 0.6832617676515272224 & 0.311021502214300052 \\
       \hline
                    & $\langle r_{eN} \rangle$ & $\langle r_{ee} \rangle$ & $\langle r^{2}_{eN} \rangle$ & $\langle r^{2}_{ee} \rangle$ \\
      \hline
             Mu$^{-}$ & 2.7271829808562793277 & 4.4392800894508588984 & 12.074193967756014098 & 25.51453633212122043 \\

 ${}^{\infty}$H$^{-}$ & 2.7101782784444203653 & 4.4126944979917277211 & 11.913699678051262274 & 25.202025291240331897 \\
        \hline
                     & $\langle r^{3}_{eN} \rangle$ & $\langle r^{3}_{ee} \rangle$ & $\langle r^{4}_{eN} \rangle$ & $\langle r^{4}_{ee} \rangle$ \\
      \hline
             Mu$^{-}$ & 77.63368957070895012 & 184.07731257386858773 & 663.917821205619807 & 1632.23450379046995 \\

 ${}^{\infty}$H$^{-}$ & 76.02309704902717911 & 180.60560023017477483 & 645.144542412219375 & 1590.09460393948530 \\
      \hline\hline
                      & $\langle r^{6}_{eN} \rangle$ & $\langle r^{6}_{ee} \rangle$ & $\langle r^{8}_{eN} \rangle$ & $\langle r^{8}_{ee} \rangle$ \\
      \hline
             Mu$^{-}$ & 9.1202102538542393$\cdot 10^{4}$ & 2.2148515807730636$\cdot 10^{5}$ & 2.33859300715081$\cdot 10^{7}$ & 5.527178579001105$\cdot 10^{7}$ \\

 ${}^{\infty}$H$^{-}$ & 8.7266142406959270$\cdot 10^{4}$ & 2.1253344237081047$\cdot 10^{5}$ & 2.20357186954914$\cdot 10^{7}$ & 5.221867644754489$\cdot 10^{7}$ \\
      \hline\hline
                      & $\langle [r_{32} r_{31}]^{-1} \rangle$ & $\langle [r_{eN} r_{ee}]^{-1} \rangle$ & $\langle [r_{32} r_{31} r_{21}]^{-1} \rangle$ & $\langle \delta({\bf r}_{eeN}) \rangle$ \\
     \hline 
             Mu$^{-}$ & 0.37822708480340377 & 0.25017716867248448 & 0.19728255748264837 & 4.877792625$\cdot 10^{-3}$ \\ 

 ${}^{\infty}$H$^{-}$ & 0.38262789034020545 & 0.25307756706456687 & 0.20082343962918944 & 5.129778775$\cdot 10^{-3}$ \\
     \hline\hline
            & $\langle \delta({\bf r}_{eN}) \rangle$ & $\nu_{eN}^{(a)}$ & $\langle \delta({\bf r}_{eN}) \rangle$ & $\nu_{ee}^{(a)}$ \\
      \hline
             Mu$^{-}$ & 0.1621506817228056 & -0.995186945189023 & 2.681680526352034$\cdot 10^{-3}$ & 0.499999989504 \\

 ${}^{\infty}$H$^{-}$ & 0.1645528728473590 & -1.00000000001778  & 2.737992126104611$\cdot 10^{-3}$ & 0.500000002446 \\
     \hline
         &  $\tau_{eN}$ & $\tau_{ee}$ & $\langle f \rangle$ & $\langle {\bf r}_{31} \cdot {\bf r}_{32} / r^3_{31} \rangle$ \\
      \hline
             Mu$^{-}$ & 0.6492013693363036596 & -0.1038138782222772366 & 0.048647215112582520641 & -0.4597494719447642817 \\

 ${}^{\infty}$H$^{-}$ & 0.6498715811920881669 & -0.1051476935659779011 & 0.048648867204549608205 & -0.4642618530806317014 \\
     \hline\hline 
     & $\langle -\frac12 \nabla^2_e \rangle$ & $\langle -\frac12 \nabla^2_N \rangle$ & $\langle \nabla_e \cdot \nabla_e \rangle$ & $\langle \nabla_e \cdot \nabla_N \rangle$ \\
      \hline
             Mu$^{-}$ & 0.2611868445353802015 & 0.554370044808132862 & 0.0319963557373724590 & -0.554370044808132862 \\

 ${}^{\infty}$H$^{-}$ & 0.2638755082721885983 & 0.560630798396681918 & 0.0328797818523047217 & -0.560630798396681918 \\
    \hline \hline
  \end{tabular}}
  \end{center}
 ${}^{(a)}$The expected cusp values (in $a.u.$) for the ${}^{\infty}$H$^{-}$ ion are $\nu_{eN} = -1.0$ and $\nu_{ee} = 0.5$. For the Mu$^{-}$ ion these 
   expected cusp values (in $a.u.$) are $\nu_{eN}$ = -0.9951869451890226269841682 (with our muon mass) and $\nu_{ee} = 0.5$. 
  \end{table}
%
 \begin{table}[tbp]
   \caption{The expectation values of some bound state properties in atomic units of the MuPs system ($\mu^{+} e^{-}_2 e^{+}$). The notation $\mu$
            designates the positively charged muon, while the notations $e^{-}$ and $e^{+}$ denote the electron and positron, respectively. $N$ is total 
            number of basis functions used.} 
     \begin{center}
    \scalebox{0.90}{%
     \begin{tabular}{| c | c | c | c | c | c | c | c |}
       \hline\hline          
  $N$ & $E$ & $\langle r^{-2}_{\mu - e^{+}} \rangle$ &  $\langle r^{-1}_{\mu - e^{+}} \rangle$ & $\langle r_{\mu - e^{+}} \rangle$ & $\langle r^{2}_{\mu - e^{+}} \rangle$ & $\langle r^{3}_{\mu - e^{+}} \rangle$ & $\langle r^{4}_{\mu - e^{+}} \rangle$ \\
      \hline
 1000 & -0.78631714258 & 0.1708472 & 0.3460478 & 3.678250 & 16.41087 & 86.4025 & 527.328 \\

 1200 & -0.78631714533 & 0.1708472 & 0.3460478 & 3.678250 & 16.41087 & 86.4025 & 527.325 \\ 
     \hline\hline
 $N$ & $\langle \frac12 p^2_{e^{-}} \rangle$ & $\langle r^{-2}_{e^{-} - e^{+}} \rangle$ &  $\langle r^{-1}_{e^{-} - e^{+}} \rangle$ & $\langle r_{e^{-} - e^{+}} \rangle$ & $\langle r^{2}_{e^{-} - e^{+}} \rangle$ & $\langle r^{3}_{e^{-} - e^{+}} \rangle$ & $\langle r^{4}_{e^{-} - e^{+}} \rangle$ \\
       \hline
 1000 & 0.323384673 & 0.3485151 & 0.4179003 & 3.488261 & 15.66631 & 85.1148 & 539.825 \\

 1200 & 0.323384675 & 0.3485152 & 0.4179000 & 3.488261 & 15.66630 & 85.1148 & 539.825 \\
     \hline\hline
  $N$ & $\langle \frac12 p^2_{e^{+}} \rangle$ & $\langle r^{-2}_{e^{-} - \mu} \rangle$ &  $\langle r^{-1}_{e^{-} - \mu} \rangle$ & $\langle r_{e^{-} - \mu} \rangle$ & $\langle r^{2}_{e^{-} - \mu} \rangle$ & $\langle r^{3}_{e^{-} - \mu} \rangle$ & $\langle r^{4}_{e^{-} - \mu} \rangle$ \\
       \hline
 1000 & 0.136686711 & 1.19458809 & 0.7257295 & 2.326014 & 7.917324 & 35.9569 & 204.744 \\

 1200 & 0.136686710 & 1.19458811 & 0.7257295 & 2.326014 & 7.917324 & 35.9569 & 204.744 \\
     \hline\hline
  $N$ & $\langle \frac12 p^2_{\mu} \rangle$ & $\langle r^{-2}_{e^{-} - e^{-}} \rangle$ &  $\langle r^{-1}_{e^{-} - e^{-}} \rangle$ & $\langle r_{e^{-} - e^{-}} \rangle$ & $\langle r^{2}_{e^{-} - e^{-}} \rangle$ & $\langle r^{3}_{e^{-} - e^{-}} \rangle$ & $\langle r^{4}_{e^{-} - e^{-}} \rangle$ \\
       \hline
 1000 & 0.591695472 & 0.2115948 & 0.36857646 & 3.594566 & 16.05697 & 86.0567 & 541.012 \\ 

 1200 & 0.591695470 & 0.2115948 & 0.36857646 & 3.594566 & 16.05697 & 86.0567 & 541.012 \\ 
     \hline\hline
 $N$ & $\langle \delta({\bf r}_{e^{-} - e^{+}}) \rangle^{(a)}$ & $\langle \delta({\bf r}_{\mu - e^{+}}) \rangle^{(a)}$ & $\langle \delta({\bf r}_{\mu - e^{-}}) \rangle^{(a)}$ & $\langle \delta({\bf r}_{e^{-} - e^{-}}) \rangle$ & 
       $\langle \delta({\bf r}_{\mu e^{-} e^{+}}) \rangle$ & 
       $\langle \delta({\bf r}_{e^{-} e^{-} e^{+}}) \rangle$ & $\langle \delta({\bf r}_{\mu e^{+} e^{-} e^{-}}) \rangle$ \\
       \hline
 1000 & 0.02441854 & 0.00162037 & 0.17431747 & 0.00430157 & 8.53207$\cdot 10^{-4}$ & 3.66758$\cdot 10^{-4}$ & 1.8176$\cdot 10^{-4}$ \\

 1200 & 0.02441854 & 0.00162037 & 0.17431728 & 0.00430159 & 8.53185$\cdot 10^{-4}$ & 3.67647$\cdot 10^{-4}$ & 1.8176$\cdot 10^{-4}$ \\
      \hline\hline
  \end{tabular}}
  \end{center}
 ${}^{(a)}$In the main text these expectation values are designated by the notations $\langle \delta({\bf r}_{+-}) \rangle, \langle \delta({\bf r}_{+ \mu}) \rangle$ and $\langle \delta({\bf r}_{- \mu}) \rangle$,
   respectively. 
   \end{table}
%
 \begin{table}[tbp]
   \caption{The expectation values of a number of bound state properties (in atomic units) of the $p \mu e_2, d \mu e_2$ and $t \mu e_2$ four-body systems. 
            The notation $\mu$ designates the positively charged muon, while the notation $a$ stands for the protium, deuterium and tritium, respectively. 
            The notation $e^{-}$ denotes the electron.} 
     \begin{center}
    \scalebox{0.90}{%
     \begin{tabular}{| c | c | c | c | c | c | c | c |}
       \hline\hline          
  system & $E$ & $\langle r^{-2}_{\mu - e^{-}} \rangle$ &  $\langle r^{-1}_{\mu - e^{-}} \rangle$ & $\langle r_{\mu - e^{-}} \rangle$ & $\langle r^{2}_{\mu - e^{-}} \rangle$ & $\langle r^{3}_{\mu - e^{-}} \rangle$ & $\langle r^{4}_{\mu - e^{-}} \rangle$ \\
      \hline
 $p \mu e_2$  & -1.150054535 & 1.531686 & 0.886755 & 1.609459 & 3.29137 & 8.1019 & 23.300 \\

 $d \mu e_2$  & -1.150704304 & 1.532542 & 0.887313 & 1.607818 & 3.28330 & 8.0659 & 23.131 \\

 $t \mu e_2$  & -1.150883319 & 1.532220 & 0.887597 & 1.607324 & 3.28200 & 8.0655 & 23.151 \\
     \hline\hline
 $N$ & $\langle \frac12 p^2_{e^{-}} \rangle$ & $\langle r^{-2}_{e^{-} - a^{+}} \rangle$ &  $\langle r^{-1}_{e^{-} - a^{+}} \rangle$ & $\langle r_{e^{-} - a^{+}} \rangle$ & $\langle r^{2}_{e^{-} - a^{+}} \rangle$ & $\langle r^{3}_{e^{-} - a^{+}} \rangle$ & $\langle r^{4}_{e^{-} - a^{+}} \rangle$ \\
       \hline
 $p \mu e_2$  & 0.568945309 & 1.543128 & 0.889721 & 1.604998 & 3.27416 & 8.0409 & 23.071 \\

 $d \mu e_2$  & 0.569517734 & 1.544663 & 0.890487 & 1.603101 & 3.26538 & 8.0038 & 22.904 \\

 $t \mu e_2$  & 0.569758745 & 1.545693 & 0.890851 & 1.602424 & 3.26298 & 7.9972 & 22.892 \\
     \hline\hline
  $N$ & $\langle \frac12 p^2_{\mu} \rangle$ & $\langle r^{-2}_{a^{+} - \mu} \rangle$ &  $\langle r^{-1}_{a^{+} - \mu} \rangle$ & $\langle r_{a^{+} - \mu} \rangle$ & $\langle r^{2}_{a^{+} - \mu} \rangle$ & $\langle r^{3}_{a^{+} - \mu} \rangle$ & $\langle r^{4}_{a^{+} - \mu} \rangle$ \\
       \hline
 $p \mu e_2$  & 2.408514781 & 0.480664 & 0.682547 & 1.509747 & 2.34618 & 3.7491 & 6.154 \\

 $d \mu e_2$  & 2.455562210 & 0.481420 & 0.683319 & 1.507081 & 2.33651 & 3.7228 & 6.090 \\

 $t \mu e_2$  & 2.468316240 & 0.482131 & 0.683880 & 1.505687 & 2.33209 & 3.7122 & 0.607 \\ 
     \hline\hline
  $N$ & $\langle \frac12 p^2_{a^{+}} \rangle$ & $\langle r^{-2}_{e^{-} - e^{-}} \rangle$ &  $\langle r^{-1}_{e^{-} - e^{-}} \rangle$ & $\langle r_{e^{-} - e^{-}} \rangle$ & $\langle r^{2}_{e^{-} - e^{-}} \rangle$ & $\langle r^{3}_{e^{-} - e^{-}} \rangle$ & $\langle r^{4}_{e^{-} - e^{-}} \rangle$ \\
       \hline
 $p \mu e_2$  & 2.428621303 & 0.488805 & 0.569490 & 2.241373 & 6.01955 & 18.7208 & 65.961 \\ 

 $d \mu e_2$  & 2.477319166 & 0.489587 & 0.569990 & 2.239114 & 6.00633 & 18.6525 & 65.601 \\ 

 $t \mu e_2$  & 2.490270302 & 0.490284 & 0.570225 & 2.238367 & 6.00299 & 18.6410 & 65.570 \\ 
     \hline\hline
 $N$ & $\langle \delta({\bf r}_{e^{-} - \mu}) \rangle$ & $\langle \delta({\bf r}_{\mu - a^{+}}) \rangle$ & $\langle \delta({\bf r}_{a^{+} - e^{-}}) \rangle$ &  
       $\langle \delta({\bf r}_{e^{-} e^{-}}) \rangle$ & $\langle \delta({\bf r}_{a^{+} e^{-} e^{-}}) \rangle$ & $\langle \delta({\bf r}_{\mu e^{-} e^{-}}) \rangle$ &  $\langle \delta({\bf r}_{\mu a^{+} e^{-}}) \rangle$ \\
       \hline
 $p \mu e_2$  & 0.21307 & 4.7078$\cdot 10^{-6}$ & 0.21489 & 0.01598 & 0.017084 & 0.017109 & 2.393$\cdot 10^{-5}$ \\

 $d \mu e_2$  & 0.21139 & 5.3055$\cdot 10^{-6}$ & 0.21318 & 0.01604 & 0.016533 & 0.016479 & 2.233$\cdot 10^{-5}$ \\ 

 $t \mu e_2$  & 0.20881 & 6.9770$\cdot 10^{-6}$ & 0.21358 & 0.01647 & 0.016308 & 0.016300 & 2.157$\cdot 10^{-5}$ \\
      \hline\hline
  \end{tabular}}
  \end{center}
   \end{table}
\end{document}